\documentclass[preprint,12pt]{elsarticle}
\usepackage{latexsym, graphicx, epsfig, amsmath, amssymb, amsfonts}
\usepackage{amsmath}
\usepackage{amssymb}
\usepackage{graphicx}
\usepackage{subfig}
\usepackage{tabu}
\usepackage{adjustbox}
\usepackage{bbold}
\usepackage{bm}
\usepackage{textcomp}
\usepackage[thinc]{esdiff}
\usepackage[dvipsnames]{xcolor}
\usepackage{comment}
\usepackage{textgreek}
\usepackage{caption}
\usepackage{lineno}

\usepackage{siunitx}
\usepackage{xcolor}
\usepackage{booktabs}
\usepackage{caption}
\usepackage{float}
\usepackage{color,soul}

\journal{Journal}
\begin{document}
\begin{frontmatter}
\title{G2\textPhi net: Relating Genotype and Biomechanical Phenotype of Tissues with Deep Learning}

\author{Enrui Zhang\textsuperscript{a}}
\author{Bart Spronck\textsuperscript{b}}
\author{Jay D. Humphrey\textsuperscript{c}\corref{cor1}}
\author{George Em Karniadakis\textsuperscript{a,d}\corref{cor1}}
\cortext[cor1]{Corresponding authors. E-mail: jay.humphrey@yale.edu (J.D.H.); george\_karniadakis@brown.edu (G.E.K.)}
\address[a]{Division of Applied Mathematics, Brown University, Providence, RI 02912, USA}
\address[b]{Department of BioMedical Engineering, Maastricht University, the Netherlands}
\address[c]{Department of Biomedical Engineering, Yale University, New Haven, CT 06520, USA}
\address[d]{School of Engineering, Brown University, Providence, RI 02912, USA}

\begin{abstract}

Many genetic mutations adversely affect the structure and function of load-bearing soft tissues, with clinical sequelae often responsible for disability or death. Parallel advances in genetics and histomechanical characterization provide significant insight into these conditions, but there remains a pressing need to integrate such information. We present a novel genotype-to-biomechanical-phenotype neural network (G2\textPhi net) for characterizing and classifying biomechanical properties of soft tissues, which serve as important functional readouts of tissue health or disease. We illustrate the utility of our approach by inferring the nonlinear, genotype-dependent constitutive behavior of the aorta for four mouse models involving defects or deficiencies in extracellular constituents. We show that G2\textPhi net can infer the biomechanical response while simultaneously ascribing the associated genotype correctly by utilizing limited, noisy, and unstructured experimental data. More broadly, G2\textPhi net provides a powerful method and a paradigm shift for correlating genotype and biomechanical phenotype quantitatively, promising a better understanding of their interplay in biological tissues.
\end{abstract}


\end{frontmatter}



\newpage
\section{Introduction}

Advances in genomics and medical genetics continue to uncover mutations that adversely affect the vasculature, with prominent examples including Marfan syndrome, vascular Ehlers-Danlos syndrome, and William’s syndrome. In many of these cases, mutation-related changes in vascular composition and biomechanical properties play key roles in both disease initiation and progression. Consequently, considerable attention continues to be devoted to comparing histomechanical properties of vessels from affected humans and animal models against those of age- and sex-matched healthy controls. Such information can provide insight into these diseases and their overall consequences on the cardiovascular system. 

Mouse models have emerged as particularly important in the study of genetically triggered vascular diseases for multiple reasons, including the now routine genetic manipulations in mice as well as their short gestational period, the availability of antibodies for biological assays, and the feasibility of miniaturized instrumentation for both \textit{in vivo} and \textit{ex vivo} assessments. Among others, we developed custom computer-controlled devices for biomechanically phenotyping murine arteries~\cite{gleason2004multiaxial,bersi2016novel} and identified protocols that ensure robust parameter estimations~\cite{ferruzzi2013biomechanical,rego2021uncertainty,bellini2014microstructurally}. Findings have revealed, for example, graduated decreases in elastic energy storage capacity in cases of increasingly severe elastopathies and progressive increases in circumferential material stiffness in enlarging thoracic aortic aneurysms~\cite{humphrey2019central,bellini2017comparison}. Although microstructurally motivated, existing constitutive relations based on continuum biomechanics are phenomenological~\cite{humphrey2003continuum}. These models cannot directly relate the mechanical behavior with either the genotype or the precise microstructure of the arterial walls. They similarly cannot delineate or predict contributions of the myriad proteins, glycoproteins, and glycosaminoglycans that constitute the arterial wall in health and disease, and cannot characterize the genotype that determines the constituents of the wall and associated biomechanical properties. With measurements of genotypical and microstructural features abundantly available through advanced experimentation, there is an unprecedented opportunity to develop novel approaches to capture better the relationship between genotype and biomechanical phenotype and to understand further the interplay between biomechanical properties and genotype and/or microstructural characteristics.


Recently, deep learning algorithms have been employed extensively in data-driven studies of mechanical behavior, ranging from engineering materials to biological tissues. Some studies focus on (constitutive-) model-based approaches, where the deep learning algorithm seeks to identify optimal material parameters in an analytically expressed constitutive model \cite{holzapfel2021predictive,guo2021cpinet,flaschel2021unsupervised,liu2019estimation,zhang2020physics,zhang2022analyses,hxj2021,liu2021knowledge,liu2020machine,yin2021non}. By utilizing preexisting constitutive models, such approaches have successfully characterized material parameters in many problems. Nevertheless, such utilization limits the dimensionality of the search space of the material behavior, hence affecting the approximation capability of the model. In view of this, other studies adopt model-free approaches, where one seeks to identify a functional form for the constitutive relation directly from data \cite{huang2020learning,linka2021constitutive,wang2021deep,liu2020generic,guo2021learning,qu2021towards,mozaffar2019deep,masi2021thermodynamics,masi2021thermodynamics2,fuhg2021physics,trask2022unsupervised}. Model-free deep learning algorithms can often capture the mechanical properties of a material more accurately, but suffer from learning that can disobey physical principles. To address this issue, namely, identifying appropriate basic assumptions for the material behavior and respecting fundamental physical principles (e.g., material symmetry, material stability, and thermodynamic consistency), some studies inject physical principles into the deep learning algorithm to limit the search to a physically admissible space \cite{masi2021thermodynamics,masi2021thermodynamics2,liu2020generic,fuhg2021physics}.

Yet, for model-free approaches that deep learn a constitutive relation, there exists an inevitable and crucial but less addressed issue. Most existing studies have sought to capture the mechanical behavior of \textit{one sample from a material}. Conversely, in the classical approach before the deep learning era, one sought to analytically identify a robust constitutive relation for \textit{multiple samples for one type of material} under specific conditions of interest (e.g., temperature), noting that the class-behavior is specified by a functional expression while the sample-behavior is described by the material parameters. In other words, practically applicable constitutive models are those that are not only capable of fitting the mechanical behavior of existing samples, but all samples from the same material (or class) as well. From a deep learning perspective, we seek a constitutive relation that can be transferred and generalized across specimens within a material class. To this end, a recent study \cite{liu2020generic} has shown that a single neural network can fit the mechanical properties for multiple samples from a class of material. However, predicting the material response for a sample at a certain deformation state still requires considerable data at neighboring deformation states, which demonstrates a weak generalizability of the learned constitutive model. Further studies are needed for deep learning realistic constitutive models that are comparable in terms of generalizability to those built analytically, which are not subject-specific.

\begin{figure}[!ht]
  \centering
  \includegraphics[width=\textwidth]{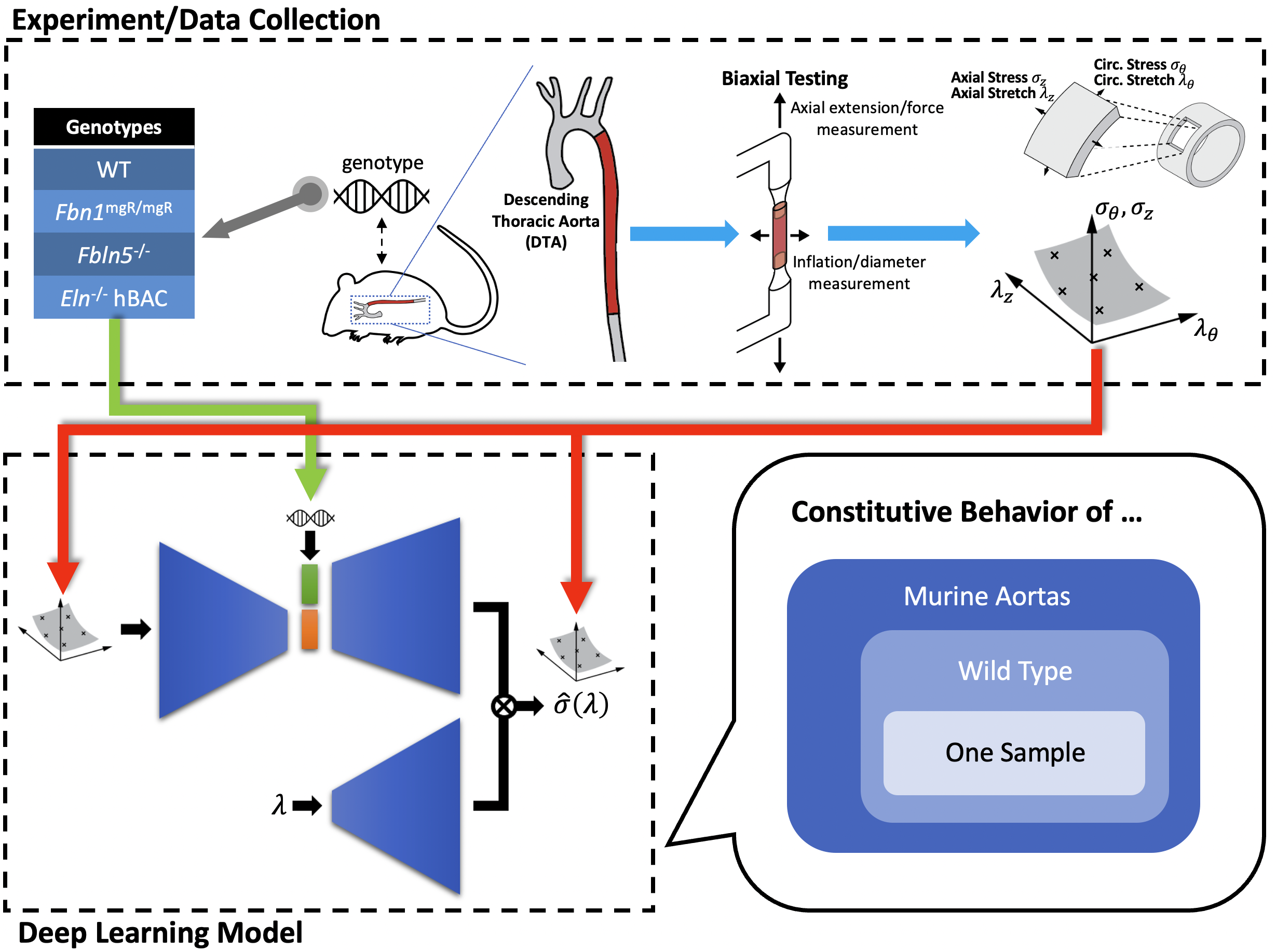}
  \caption{\textbf{Overview of the Workflow in G2\textPhi net with Experimental Measurements Informing the Deep Learning Model.} We use biaxial tests to measure the biomechanical properties of the descending thoracic aorta (DTA) from mice with different mutations that affect the elastic fibers. Through the experimental measurements, circumferential ($\sigma_{\theta}$) and axial ($\sigma_z$) stress are determined for different combinations of circumferential ($\lambda_{\theta}$) and axial ($\lambda_z$) stretch. We then use the experimental results to train our deep learning model to capture the aortic constitutive relation in a hierarchical way: general mechanical behavior across multiple genotypes, behavior for a certain genotype, and behavior for a certain sample of a certain genotype. We view genotypes as material classes that possess distinct biomechanical properties. Illustrative data are shown in Fig.~\ref{fig:experiment}. The detailed architecture of our deep learning model is explained in Fig.~\ref{fig:model}.}
  \label{fig:abstract}
\end{figure}

In this paper, we propose a genotype-to-biomechanical-phenotype neural network (G2\textPhi net), a generalizable, model-free neural operator architecture for capturing the biomechanical properties of soft tissues and identifying correlations with genetic mutations. To demonstrate the utility of G2\textPhi net, we seek to model biaxial mechanical data for the descending thoracic aortas (DTAs) from four different genotypes of mice related to elastopathies. The workflow is summarized in Fig.~\ref{fig:abstract}. We firstly collect data characterizing the biomechanical behavior, then construct and train the G2\textPhi net with the biomechanical data and genotype. Notably, we view the genotype as material classes that possess distinct biomechanical properties. The trained network captures genotype-specific constitutive relations at three levels (Fig.~\ref{fig:abstract}): murine aortas in general, aortas from a particular genotype, and a specific sample of the aortas. In the following, we first summarize the biaxial tests and the architecture of G2\textPhi net, and then present results of G2\textPhi net, followed by a discussion.

\section{Results}

\subsection{Biomechanical Data}

\begin{figure}[!ht]
  \centering
  \includegraphics[width=\textwidth]{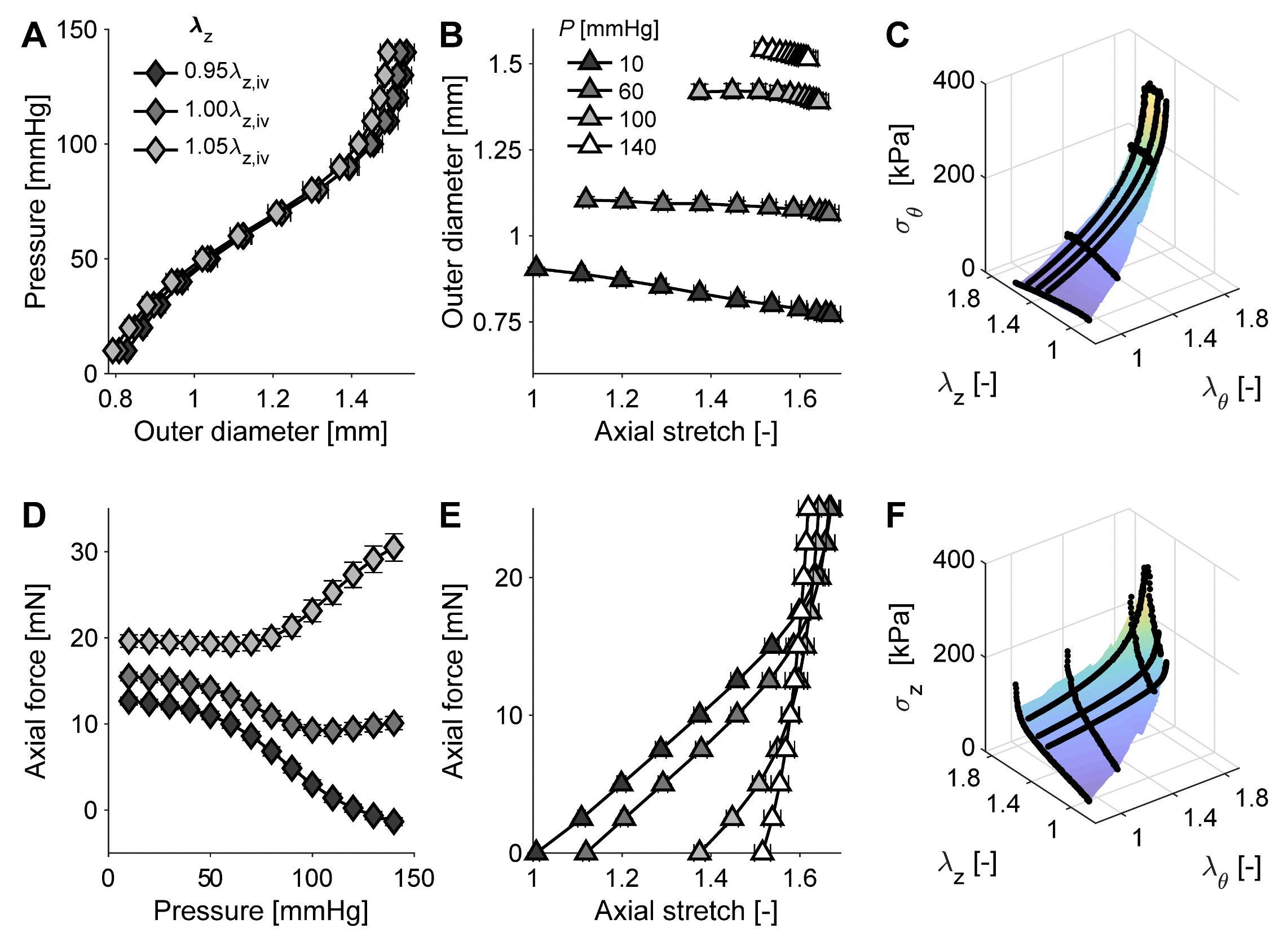}
  \caption{ \textbf{Illustration of Experimental Biomechanical Data Used for Deep Learning.} (A,D) Data from distension experiments at three constant levels of axial stretch ($\lambda_z$), relative to the in vivo value of axial stretch $\lambda_{z\text{,iv}}$. (B,E) Data from axial extension experiments at four constant pressures ($P$). Symbols and whiskers indicate means $\pm$ standard errors across these 8 wild-type (WT) control samples. (C,F) Circumferential ($\sigma_{\theta}$, C) and axial ($\sigma_z$, F) stress as a function of $\lambda_{\theta}$ and $\lambda_z$ for one representative WT sample.}
  \label{fig:experiment}
\end{figure}

The testing procedure is summarized in the top panel of Fig.~\ref{fig:abstract}. Biaxial (circumferential and axial) mechanical data were collected for DTAs from 28 mice from four different genotypes: wild-type (WT, 8 samples), $\textit{Fbn1}^{\text{mgR/mgR}}$ (8 samples), $\textit{Fbln5}^{\text{-/-}}$ (5 samples), and $\textit{Eln}^{\text{-/-}}\text{ hBAC}$ (7 samples; see Materials and Methods for experimental details). These genotypes are known knock-out mouse models of interest that change structural constituents of the arterial wall, hence resulting in distinct biomechanical properties. In the biomechanical testing, briefly, arteries were mounted between glass micropipets, mechanically preconditioned, then tested using a seven-step protocol consisting of (1) three distension cycles from $10$ to $140 \, \text{mmHg}$ at an axial stretch corresponding to 95\%, 100\%, and 105\% of the \textit{in vivo} value, and (2) four axial extension cycles at constant pressures of $10$, $60$, $100$, and $140 \, \text{mmHg}$. A representative experimental data set is shown in Figs~\ref{fig:experiment}A, B, D, and E. Given the geometry and applied loads, circumferential ($\sigma_{\theta}$) and axial ($\sigma_z$) Cauchy stress are calculated as visualized in Figs~\ref{fig:experiment}C and F as a function of both circumferential ($\lambda_{\theta}$) and axial ($\lambda_z$) stretch (see \cite{ferruzzi2013biomechanical} for detailed methods). Complete data for all four genotypes are included in Section S1 (Fig. S1) in Supporting Information (SI).

Because the stress-stretch data distribute nonuniformly and are distinct for different specimens within each group, we preprocessed these data so that the stress $(\sigma_z,\sigma_\theta)$ is available on a fixed set of $m\times m$ ($m=31$ herein) uniform, structured stretch states in the domain $(\lambda_z,\lambda_\theta)\in[1.00,1.65]^2$. This preprocessing is accomplished by fitting the raw data consistent with physical principles (e.g., convexity) and reasonable assumptions (nonlinear and anisotropic) regarding the shape of the stress-stretch relationship, which is explained in Materials and Methods. Due to significant changes of stress over the stretch domain (over two orders of magnitude), we define a log-transformed, normalized value of stress $\tilde{\sigma}_i$ ($i=z,\theta$) according to
\begin{equation}
    \label{eqn:stress_norm}
    \tilde{\sigma}_i=\ln(\frac{\sigma_i}{\sigma_0}+1.0),
\end{equation}
where $\sigma_0=1 \, \text{kPa}$. In the following sections, unless otherwise noted, we simplify the notation of $\tilde{\sigma}$ to be $\sigma$ to avoid multiple accents.

\subsection{Genotype-to-Biomechanical-Phenotype Neural Network (G2\textPhi net)}

\begin{figure}[!ht]
  \centering
  \includegraphics[width=\textwidth]{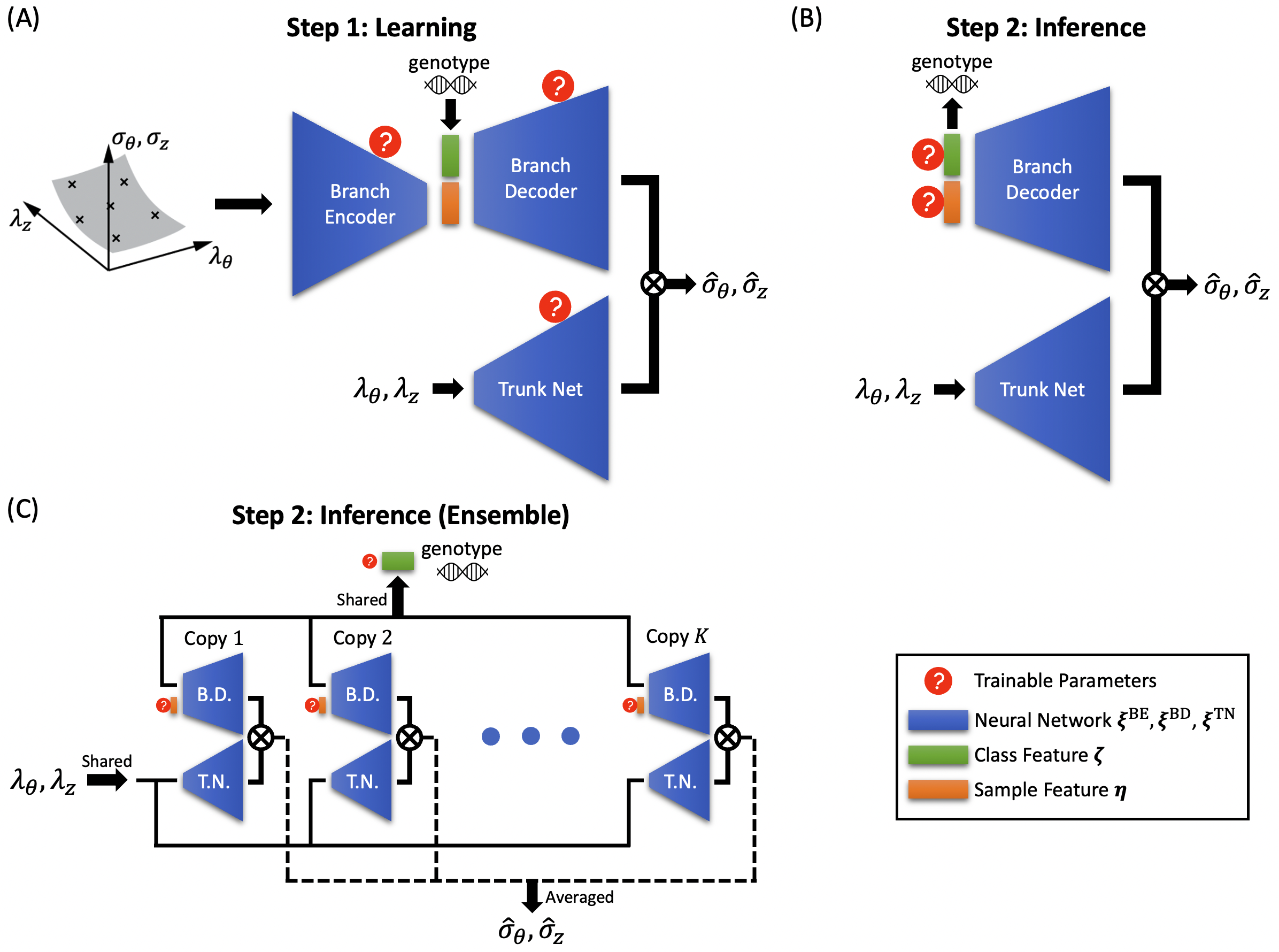}
  \caption{\textbf{Architecture of G2\textPhi net.} (A) The learning stage. For each aortic sample, we input stress-stretch measurements into the branch encoder, which compresses the input into the sample feature $\bm{\eta}$ in the latent space. The genotype (class feature $\bm{\zeta}$) serves as additional input to the latent space. The branch encoder takes the combination of class and sample features as inputs. The trunk net takes an arbitrary stretch state $(\lambda_\theta,\lambda_z)$ as inputs. The inner product of the outputs from the branch decoder and the trunk net is the predicted stress $(\hat{\sigma}_\theta,\hat{\sigma}_z)$ at a given stretch state $(\lambda_\theta,\lambda_z)$. The trainable parameters for this stage include the weights and biases of the three sub-networks. (B) The inference stage. The branch decoder and trunk net serve together as an approximation of the stress-stretch relationship parameterized by the class feature $\bm{\zeta}$ (genotype) and sample feature $\bm{\eta}$. The weights and biases of the two sub-networks are fixed to be their values upon completion of the learning stage, while the class and sample features are trainable. (C) The inference stage using ensemble. With $K$ copies of the trained model, their branch decoders (B.D.) and trunk nets (T.N.) are integrated into a single mega-model for inferring new samples. Input to the trunk net $(\lambda_\theta,\lambda_z)$ and class feature $\bm{\zeta}$ are shared across model copies. The stress prediction $(\hat{\sigma}_\theta,\hat{\sigma}_z)$ is the mean value across $K$ copies.}
  \label{fig:model}
\end{figure}

Inspired by the encoder-decoder~\cite{badrinarayanan2017segnet} and DeepONet~\cite{lu2021learning,yin2022simulating,goswami2021physics,yin2022interfacing} architectures, we design a new neural network architecture, G2\textPhi net, for characterization and classification of soft biological tissues involving genotype and biomechanical phenotype. G2\textPhi net comprises three subnetworks (Fig.~\ref{fig:model}): a branch encoder (with trainable parameters $\bm{\xi}^\text{BE}$), a branch decoder ($\bm{\xi}^\text{BD}$), and a trunk net ($\bm{\xi}^\text{TN}$), where naming of the branch and trunk is inherited from the original DeepONet paper. We adopt a two-step strategy to use G2\textPhi net. The first step is the \textit{learning stage} (Fig.~\ref{fig:model}A), in which G2\textPhi net seeks to capture the genotype-dependent stress-stretch relation through training. The branch encoder takes the stress-stretch curves as inputs and compresses them into the sample feature (denoted as $\bm{\eta}$, with $d_{\bm{\eta}}$ dimensions) in the latent space. This sample feature, together with genotype as the class feature (denoted as $\bm{\zeta}$, with $d_{\bm{\zeta}}$ dimensions) fed at the latent space, serves as the input of the branch decoder. The branch encoder and branch decoder work together to identify the minimal necessary parameters for describing the material properties. The trunk net, on the other hand, takes an arbitrary stretch state $(\lambda_\theta,\lambda_z)$ as input. The final output of G2\textPhi net is the stress prediction $(\hat{\sigma}_\theta,\hat{\sigma}_z)$ for stretch $(\lambda_\theta,\lambda_z)$, which is calculated from the inner product of outputs from the branch decoder and the trunk net. We train G2\textPhi net by minimizing the mismatch between the stresses from the input (of the branch encoder) and the output (of the branch decoder and trunk net). After training, the latter two subnetworks together serve as the approximation of the stress-stretch relationship:
\begin{equation}
\label{eqn:BDTN}
    (\hat{\sigma}_\theta,\hat{\sigma}_z)=\mathcal{NN}_{\bm{\xi}^\text{BD},\bm{\xi}^\text{TN}}(\lambda_\theta,\lambda_z;\bm{\zeta},\bm{\eta}),
\end{equation}
where $\bm{\xi}^\text{BD}$ and $\bm{\xi}^\text{TN}$ determine the functional form, and the class feature (genotype as material classes) $\bm{\zeta}$ and the sample feature $\bm{\eta}$ serve as ``material parameters'' of the learned constitutive relation. Note that the sample feature $\bm{\eta}$ is not mechanically or biologically interpretable due to the unsupervised nature of the encoder-decoder structure.

The second step is the \textit{inference stage} (Fig.~\ref{fig:model}B), where we use the learned stress-stretch relationship Eq.~\ref{eqn:BDTN} to infer the stress-stretch relation and genotype for new aorta samples, based on limited and scattered measurements of stress and stretch. There are two main differences regarding the setup of G2\textPhi net in this step: (1) the branch encoder is not used, and (2) the (previously) trainable parameters of the branch decoder $\bm{\xi}^\text{BD}$ and the trunk net $\bm{\xi}^\text{TN}$ are no longer trainable, while class feature $\bm{\zeta}$ and sample feature $\bm{\eta}$ are trainable. Similar to a best-fit procedure where one estimates values of material parameters from experimental measurements, G2\textPhi net looks for $\bm{\zeta}$ and $\bm{\eta}$ that match the stress-stretch measurements optimally. After this step, G2\textPhi net is expected to reconstruct the entire functional stress-stretch relation (according to Eq.~\ref{eqn:BDTN}) and simultaneously predict the source genotype (according to the updated value of $\bm{\zeta}$).

Our relatively small dataset (28 specimens across four genotypes) makes it arduous to build a suitable deep learning model, which typically demands big data. To address this issue, we introduced multiple techniques to improve model performance with limited data. The most important technique is \textit{ensemble learning}~\cite{lakshminarayanan2016simple}. We find that the different random initializations for G2\textPhi net influence its performance, especially genotype prediction. Such sensitivity is possibly caused by the nature of smallness in our problem -- small data and small dimension of the latent space. To eliminate such unwanted fluctuations of model performance, we adopted the ensemble learning method (Fig.~\ref{fig:model}C). We firstly obtain multiple copies of G2\textPhi net in the learning stage (Fig.~\ref{fig:model}A) by (1) training $K_1$ independent copies with different random seeds and (2) saving the trained models after $K_2$ different numbers of epochs in the late stage of the training process. In the inference stage, these $K=K_1K_2$ copies are integrated into a mega-model (Fig.~\ref{fig:model}C), where the class feature $\bm{\zeta}$ is shared across all copies and the sample feature $\bm{\eta}_{(i)}$ ($i\in{1,...,K}$) is independently defined per copy. The mega-model predicts the mean stress from all copies as its output:
\begin{equation}
\label{eqn:ens_mean}
    (\hat{\sigma}_\theta,\hat{\sigma}_z)=\frac{1}{K}\sum_{i=1}^{K}(\hat{\sigma}_{\theta(i)},\hat{\sigma}_{z(i)})=\frac{1}{K}\sum_{i=1}^{K}\mathcal{NN}_{\bm{\xi}_{(i)}^\text{BD},\bm{\xi}_{(i)}^\text{TN}}(\lambda_\theta,\lambda_z;\bm{\zeta},\bm{\eta}_{(i)}),
\end{equation}
where subscript $(i)$ indicates the specific copy of the model. We explain the details of G2\textPhi net including necessary techniques such as mixup regularization~\cite{zhang2017mixup} in Materials and Methods and Section S2 in SI.


\subsection{Learning Genotype-Dependent Constitutive Behavior}

We firstly train G2\textPhi net to let it capture the constitutive relationship among various genotypes (learning stage; Fig.~\ref{fig:model}A). As explained in the foregoing section, we provide information on the stress-strain relationship and genotype to G2\textPhi net, which seeks to minimize the mismatch between the input and output relationship. After the learning stage, we preliminarily test model efficacy by providing the same input information as training data from an unseen aortic specimen(s) as test data and examine the reconstruction of the stress-strain relationship. Note that the test here is not yet the inference stage for new specimens -- we still utilize the entire deep learning model including the branch encoder, and the model is still informed with complete data on $m\times m$ grid points from the stress-strain relationship through the branch encoder. To accurately assess the model performance for our target problem with limited data, we conduct $5$-fold cross validation throughout this work.

\begin{figure}[!ht]
  \centering
  \includegraphics[width=0.9\textwidth]{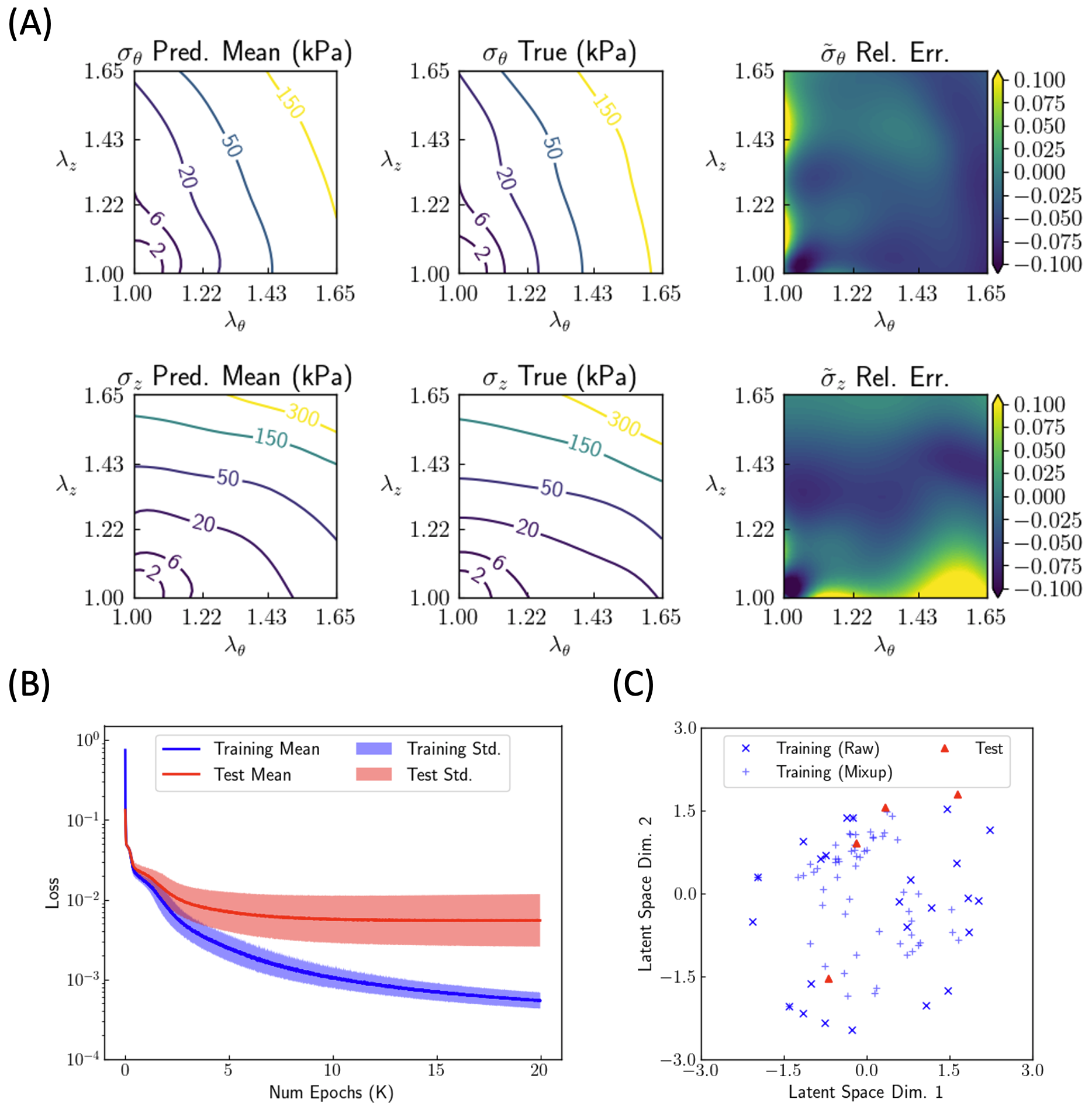}
  \caption{\textbf{Results of G2\textPhi net Training}. The dimension of sample feature is $d_{\bm\eta}=2$. (A) Reconstruction of the stress-stretch relationship ($\sigma_z$ and $\sigma_\theta$ as functions of $\lambda_z$ and $\lambda_\theta$) for a typical test sample averaged over model predictions across random seeds. Model prediction (output of the branch decoder and trunk net), ground truth value (input of the branch encoder), and their difference are shown. The predicted and true values of the stress are displayed by their respective physical values ($\sigma_i$ where $i=z,\theta$; in kPa), while their difference is displayed as the relative error of the normalized stress $\tilde{\sigma}_i$. (B) The reconstruction loss in the learning stage for training data and test data over all random seeds, all training/testing samples. Both the mean value and the standard deviation are shown. (C) Visualization of the latent space of sample information with dimension $d_{\bm\eta}=2$ for a typical train-test split.}
  \label{fig:learning}
\end{figure}

We display results in Fig.~\ref{fig:learning} for $d_{\bm{\eta}}=2$, i.e., two-dimensional sample feature. To build ensembles and examine variations in model performance caused by random initialization in G2\textPhi net, we run G2\textPhi net with 67 random seeds, which produce similar but not identical results for the reconstructed stress-strain relationship. The mean prediction of the stresses and its comparison with the true stresses for a typical test sample aorta are shown in Fig.~\ref{fig:learning}A. The predicted and true values of stress $\sigma_i$ ($i=z,\theta$) are in kPa, while their difference is displayed by the relative error of the normalized stress $\tilde{\sigma}_i$. The mean prediction of $\sigma_\theta$ and $\sigma_z$ matches well with the respective true values (with relative $L^2$ error of $\tilde{\sigma}_\theta$ and $\tilde{\sigma}_z$ being roughly $4.5\%$ and $4.4\%$, respectively), indicating that G2\textPhi net learned to reconstruct the stress-stretch relationship. We found some fluctuations, however, in prediction across different random seeds with relative magnitude around $10\%$. Such a seemingly large fluctuation is reasonable because of (i) the sample feature as a bottleneck of the encoder-decoder with merely two dimensions, and (ii) the limited number (24) of training samples. In Fig.~\ref{fig:learning}B, we show the evolution of the reconstruction loss for both training and testing data, including the mean value and standard deviation calculated for all testing data (from five-fold cross validation) and all random seeds. After training around 13K epochs, the mean test loss reached a plateau of $\mathcal{O}(10^{-3})$ and did not decrease further as training proceeded, which indicates that G2\textPhi net was sufficiently trained. We stop the training at 20K epochs to ensure that individual models are generally well trained. Fig.~\ref{fig:learning}C visualizes a two-dimensional latent space of the sample feature for a typical train-test split, where the data points refer to those from the raw training data, training data from mixup regularization, and test data. Mixup data points generally ``interpolate" the raw data points, thus the training data are more densely distributed around the origin and G2\textPhi net is anticipated to be better regularized. Test points, on the other hand, are also located roughly in the same range as the training data, indicating that G2\textPhi net should be capable of reconstructing the stress pattern for test data as accurately as for training data. With such a distributed value of sample feature, G2\textPhi net is expected to capture variations of mechanical behavior among individual samples, in addition to among different genotypes for which the class feature has been provided explicitly to the branch decoder.

\subsection{Sample Inference for Genotype Classification and Stress Reconstruction}

From the results of the learning stage in the foregoing section, we have demonstrated the capability of G2\textPhi net to capture the genotype-specific, sample-specific constitutive behavior. The actual performance can only be tested, however, by using the learned parameterized stress-strain relationship captured by the branch decoder and the trunk net (see Eq.~\ref{eqn:BDTN} and Fig.~\ref{fig:model}B) to fit new samples, where stress data $\{(\sigma_z^{(i)},\sigma_\theta^{(i)})\}_{i=1}^{m'}$ are provided only at limited stretch states $\{(\lambda_z^{(i)},\lambda_\theta^{(i)})\}_{i=1}^{m'}$ ($m'<<m^2$). In light of this, we designed two cases to evaluate the performance of the learned constitutive relation: (Setup 1: preprocessed structured data) stress is provided to the model on a $3\times 3$ structured grid of stretch states from the preprocessed data; (Setup 2: raw unstructured data) stress is provided to the model on 22 stretch states in the real experimental measurements. For both setups, as the ``material parameters'' $\bm{\zeta}$ and $\bm{\eta}$ are updated to fit the provided data, we expect the model to concurrently reconstruct the entire stress-stretch relationship and identify the source genotype. This test mimics the practical scenario, where one has a limited number of measurement points that are obtained possibly at a set of unstructured stretch states. For Setup 1, stress data sparsely cover the stretch domain $(\lambda_{\theta},\lambda_z)\in[1.00,1.65]^2$, hence we examine the performance of the reconstructed stress-stretch relationship mainly in terms of interpolation capability. In Setup 2, with the raw data available only in certain region of the stretch domain, we aim to test model performance in terms of extrapolation capability.

\begin{figure}[!ht]
  \centering
  \includegraphics[width=0.89\textwidth]{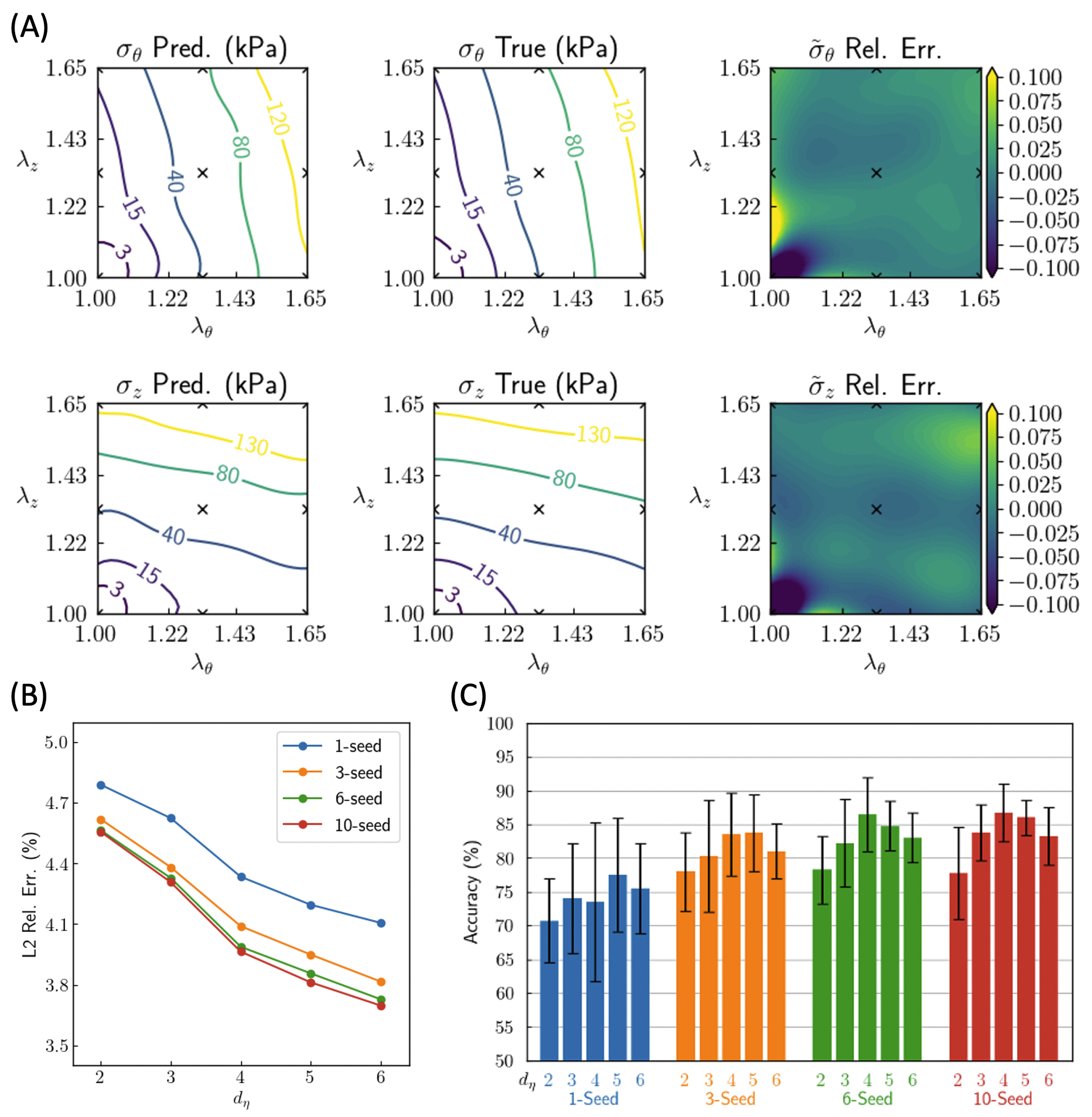}
  \caption{\textbf{Reconstruction and Classification Results of Sample Inference based on Structured Data}. Stress data are provided on $3\times 3$ uniform grid points on the stretch plane. (A) Reconstruction of the stress-stretch relationship ($\sigma_z$ and $\sigma_\theta$ as functions of $\lambda_z$ and $\lambda_\theta$) for a typical test sample from a 10-seed ensemble model. The predicted and true values of the stress are displayed by their respective physical values ($\sigma_i$ where $i=z,\theta$; in kPa), while their difference is displayed as a relative error for normalized stress $\tilde{\sigma}_i$. (B) $L_2$ relative error of the predicted normalized stress $\tilde{\sigma}_i$ with different sizes of ensemble (1-, 3-, 6- and 10-seed) and different dimensions of sample feature ($d_{\bm\eta}\in\{2,3,4,5,6\}$). Each dot is an average value over 20 runs with different choices of random seeds. (C) Classification accuracy for different sizes of ensemble and different dimensions of sample feature. Each bar is the average value over 20 runs with different choices of random seeds.}
  \label{fig:inference_S}
\end{figure}

Results for Setup 1 (preprocessed structured data) are in Fig.~\ref{fig:inference_S}. Fig.~\ref{fig:inference_S}A compares the reconstructed (predicted) and the true stress-stretch relationship for a typical test sample from a 10-seed ensemble mega-model with $d_{\bm{\eta}}=4$. The stress data are available to the model only on $3\times 3$ grid points of the stretch state as marked by x's in the figure. Notably, the measurements are placed to sparsely cover the boundary of the stretch domain $(\lambda_{\theta},\lambda_z)\in[1.00,1.65]^2$, so that reconstruction of the stress profile mainly involves interpolation instead of extrapolation. Even with such a small dataset, our trained model can accurately reconstruct the entire stress-stretch function, with a $L^2$ relative error of $1.6\%$ for $\sigma_{\theta}$ and $2.2\%$ for $\sigma_z$. In addition to results for individual test samples, we also examine the statistics across all test samples and how the model parameters influence model performance. Fig.~\ref{fig:inference_S}B shows the mean $L^2$ relative error of the normalized stress $\tilde{\sigma}_i$ ($i=z,\theta$) (out of 20 runs with different random seeds) of the predicted stress for different sizes of the ensemble as well as different values of $d_{\bm{\eta}}$ ($\in\{2,3,4,5,6\}$). The mean $L^2$ relative error is smaller for larger ensembles, indicating one advantage of building ensembles from individual models for a more accurate stress reconstruction. As the dimension of the sample feature space ($d_{\bm{\eta}}$) increases, the reconstruction error decreases, which demonstrates that a larger latent space allows more flexibility in expressing the stress-stretch relationship, hence capturing the detailed, sample-specific constitutive behavior more accurately. However, such flexibility does not always bring about benefits, as illustrated by the classification results. Fig.~\ref{fig:inference_S}C shows classification accuracy for unseen test samples for different sizes of ensemble and different dimensions of the sample feature space. The mean and standard deviation are calculated through cross-validation and different random seeds. We found that a larger ensemble not only increases the mean prediction accuracy, it also decreases the fluctuation of such accuracy among different test sets and randomization, hence improving the reliability and robustness of the deep learning model. For the classification task, the accuracy no longer monotonically increases as $d_{\bm{\eta}}$ increases. As seen in Fig.~\ref{fig:inference_S}C, the optimal dimension for the sample feature space is $d_{\bm{\eta}}=4$ or $5$. With an excessively large $d_{\bm{\eta}}$, the training data encoded into the latent space are too sparse during the learning stage, and the model is likely not to generalize well into the test data and hence performs worse in the classification task. The best performance is achieved for a 10-seed ensemble mega-model with $d_{\bm{\eta}}=4$, which yields mean classification accuracy $87\%$ and $L_2$ relative error of stress reconstruction $3.7\%$.

\begin{figure}[!ht]
  \centering
  \includegraphics[width=0.88\textwidth]{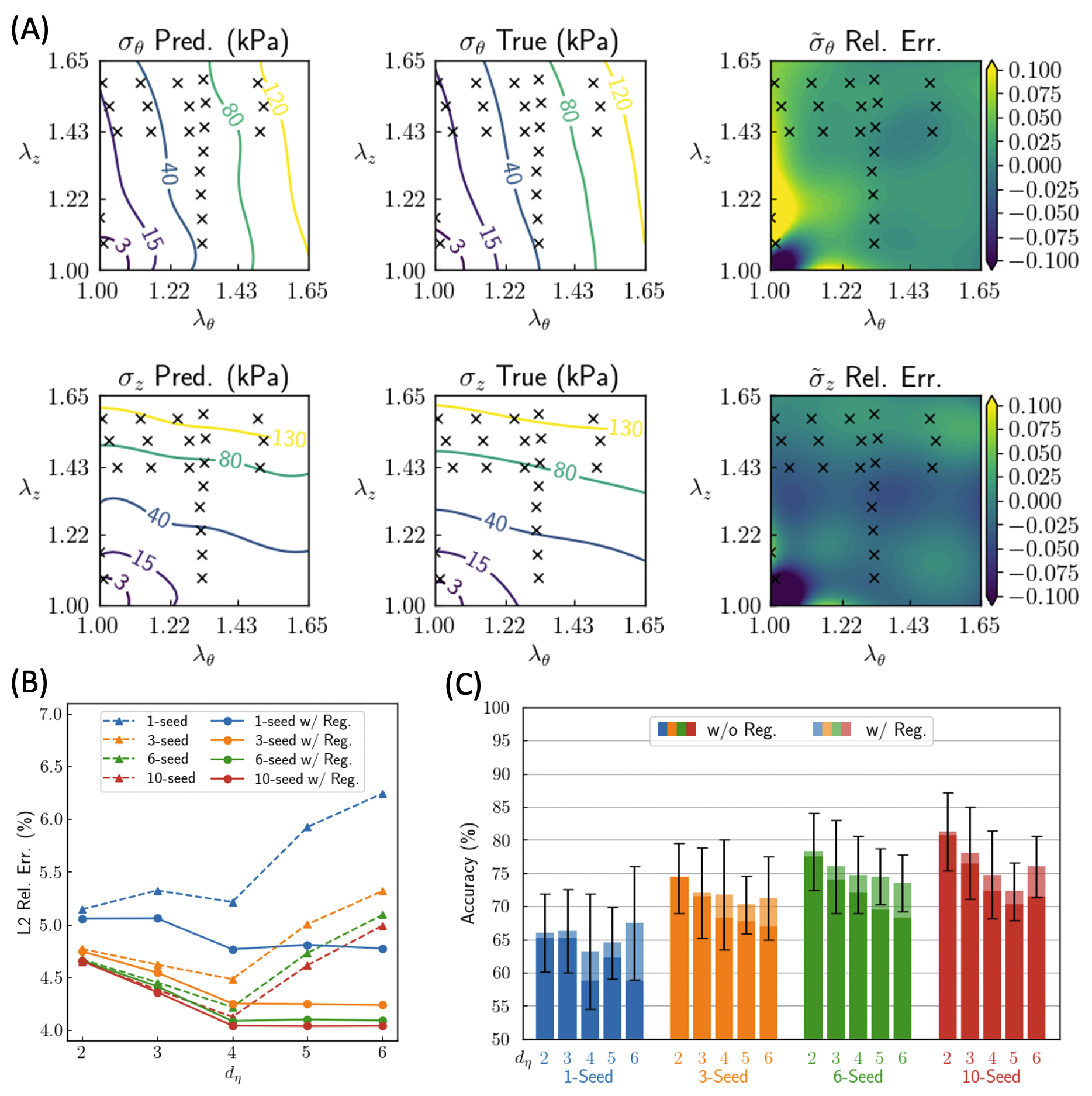}
  \caption{\textbf{Reconstruction and Classification Results of Sample Inference based on Unstructured Data}. Stress data are provided for 22 unstructured points from experimental measurements. (A) Reconstruction of the stress-stretch relationship ($\sigma_z$ and $\sigma_\theta$ as functions of $\lambda_z$ and $\lambda_\theta$) for a typical test sample for a 10-seed ensemble model. The predicted and true values of the stress are displayed by their respective physical values ($\sigma_i$ where $i=z,\theta$; in kPa), while their difference is displayed as a relative error for normalized stress $\tilde{\sigma}_i$. (B) $L_2$ relative error of the predicted normalized stress $\tilde{\sigma}_i$ with different sizes of ensemble (1-, 3-, 6- and 10-seed), different dimensions ($d_{\bm\eta}\in\{2,3,4,5,6\}$) and regularization for the sample feature (without/with regularization). Each dot is the average value over 20 runs with different choices of random seeds. (C) Classification accuracy for different sizes of ensemble, different dimensions and different regularization setups of sample feature. Each bar is the average value over 20 runs with different choices of random seeds.}
  \label{fig:inference_U}
\end{figure}

Results for Setup 2 (raw unstructured data) are in Fig.~\ref{fig:inference_U}. The stress data comes from raw experimental measurements, where they cluster around a subregion within the stretch domain $(\lambda_{\theta},\lambda_z)\in[1.00,1.65]^2$. Inevitably, reconstruction of the entire stress-stretch relationship involves extrapolation based on the available data, hence increasing the difficulty for the tasks of stress reconstruction and genotype classification. For this Setup, we subsequently provided our model with 22 stress measurements, $\sim 1.5$ times more than Setup 1. Nevertheless, the data still accounts for only $\sim 2\%$ of the original experimental measurements. Similar to Setup 1, we show the reconstruction of stress-stretch relationship for a typical test sample in a 10-seed ensemble (Fig.~\ref{fig:inference_U}A), $L_2$ relative error of the normalized stress (Fig.~\ref{fig:inference_U}B), and classification accuracy for different model parameters (Fig.~\ref{fig:inference_U}C). For the test sample shown in Fig.~\ref{fig:inference_U}A, the stress profile can be accurately reconstructed, with the $L_2$ relative error as small as $2.9\%$ for $\sigma_{\theta}$ and $2.4\%$ for $\sigma_z$. It is worth noting that the stress prediction is accurate even in regions where there are no data (e.g., $(\lambda_{\theta},\lambda_z)\in[1.40,1.65]\times[1.00,1.40]$), which demonstrates that our trained model has effectively learned the general pattern of the stress-stretch relationship for these murine aortas. In Figs~\ref{fig:inference_U}B (solid lines) and C (opaque bars, labeled with ``w/o Reg.''), however, we notice some phenomena different from Setup 1 -- the $L_2$ relative error increases as $d_{\bm{\eta}}$ increases beyond 4 (Fig.~\ref{fig:inference_U}B), while the classification accuracy decreases monotonically as $d_{\bm\eta}$ increases within $\{2,3,4,5,6\}$ (Fig.~\ref{fig:inference_U}C). We attribute such phenomena to overfitting in the stress reconstruction, which is greater for this extrapolation case. An excessively large $d_{\bm\eta}$ endows G2\textPhi net with excessive flexibility in fitting capability, so that the model seeks to fit the data slightly better at the cost of a significant drop of the overall reconstruction accuracy and classification accuracy, especially for the extrapolation regions.

To deal with the overfitting issue for Setup 2 (raw unstructured data), we modify the loss function for the sample fitting stage by including an additional regularization term  to penalize the sample feature that deviates too much from the origin (see Materials and Methods for details). The results with the modified loss function are shown in Figs. \ref{fig:inference_U}B (dashed lines) and C (translucent bars, labeled with ``w/ Reg."). This modification significantly reduces the $L_2$ relative error of stress reconstruction and increases the classification accuracy, especially for large values of $d_{\bm\eta}$. With such a technique, the best performance comes from the 10-seed ensemble with $d_{\bm\eta}=2$, which yields a mean classification accuracy of $81\%$ and $L_2$ relative error of stress reconstruction $4.0\%$.

\section{Discussion}

Continuing discoveries in genetics increasingly reveal new vascular phenotypes, many with critical clinical importance. Whereas it was long thought \cite{roach1957reason} that the passive mechanical behavior of an artery was determined by two primary structural proteins, elastin and collagen, we now know that many accessory proteins and glycoproteins as well as different matricellular proteins similarly play key roles in determining the complex nonlinear material properties. As examples, although type I collagen is the most abundant type in an artery, there are many different types of collagen within the wall. Mutations to the alpha1 helix of collagen III and collagen V molecules ($\textit{Col3a1}^{\text{+/-}}$ and $\textit{Col5a1}^{\text{+/-}}$) result in a vascular Ehlers-Danlos phenotype, namely a fragile dissection-prone artery \cite{liu1997type,segev2006structural}.
%
%
Mutations to genes that encode matricellular proteins, such as thrombospondin-2 ($\textit{Tsp2}^{\text{-/-}}$), can also increase the structural vulnerability of the arterial wall, in part by compromising collagen fibrillogenesis \cite{bellini2017hidden}. These are just a few of the many cases wherein specific mutations lead to particular types of changes in material properties within particular regions of the arterial tree. Fortunately, with available mouse models we can now begin to build consistent data bases for analysis across these many cases.

Similarly, many different classes of drugs have been developed or used to treat the diverse vascular pathologies that present clinically. Not surprisingly, many of these drugs have also been used to treat mouse models –- they include different classes of anti-hypertensive, anti-inflammatory, and lipid-lowering drugs, and those that reduce proteolytic activity to name a few. Just as there is a pressing need to discover genotype-phenotype relations for the myriad constituents that contribute to the biomechanics of the arterial wall, there is similarly a need to correlate drug treatments with the resulting phenotype, thus establishing efficacy in terms of effects on vascular structure and function rather than more general outcomes such as morbidity or mortality. In neither case can current phenomenological constitutive models provide the necessary connections.

For purposes of illustration, we focused herein on consequences of mutations to elastin and two elastin-associated glycoproteins, fibrillin-1 and fibulin-5. Elastic fibers constitute about one-third of the murine aortic wall by dry weight and endow the aorta with its resilience, a characteristic critical to its function as a conduit vessel. These fibers consist of ~90\% elastin and ~10\% elastin-associated glycoproteins, which include the fibrillins and fibulins but others as well. The clinical importance of elastic fiber integrity is revealed by multiple conditions. Williams’ syndrome results from mutations that include the elastin gene; it is characterized by supravalvular aortic stenosis (SVAS), that is, marked narrowing of the proximal aorta. Such narrowing increases the workload on the left ventricle and thus increases morbidity. William’s syndrome has been studied using different animal models, including the $\textit{Eln}^{\text{+/-}}$ mouse \cite{li1998novel}, which produces about 50\% of the normal elastin. Our two models, hBAC-mNull and hBAC-mWT (see below), bracket this model by producing 15\% and 115\% of the normal elastin, respectively \cite{hirano2007functional}. Marfan syndrome results from mutations to the gene that encodes fibrillin-1, a glycoprotein that appears to help confer the normally long half-life of elastic fibers. The vascular phenotype seen in Marfan syndrome has been modeled in mice using a $\textit{Fbn1}^{\text{C1041G/+}}$ mutation, which has dramatic effects on the mechanical properties \cite{chung2007loss}. Among the many clinical events, Marfan patients often experience aortic root and proximal aortic dilatation, often leading to dissection and/or rupture. Our choice of models, $\textit{Fbn1}^{\text{mgR/mgR}}$, results in a severe Marfan-like phenotype \cite{cook2014abnormal}. Mutations to fibulin-4 also lead to thoracic aneurysms \cite{huang2010fibulin} whereas those to fibulin-5 lead primarily to an accelerated aging phenotype and tortuosity \cite{ferruzzi2015decreased}. Vascular aging is, of course, one of the key risk factors for many cardiovascular, neurovascular, and renovascular diseases \cite{o2007mechanical,greenwald2007ageing}. 

We previously showed that the severity of the different elastopathies studied herein could be quantified, in part, via the elastic energy storage capacity. Under physiologic loading, mean values of stored energy were found to be \cite{humphrey2019central}: 156, 114, 59, 43, and 18 kPa for non-stenotic and non-aneurysmal hBAC-mWT, WT, $\textit{Fbn1}^{\text{mgR/mgR}}$, $\textit{Fbln5}^{\text{-/-}}$, and hBAC-mNull DTAs, respectively. Although these values ordered nicely, they are not diagnostic. For example, the mean energy storage was 40-60 kPa for both the non-aneurysmal $\textit{Fbln5}^{\text{-/-}}$ and an aneurysmal $\textit{Fbn1}^{\text{mgR/mgR}}$ thoracic aorta. Indeed, although the constitutive relation used for quantification has described biaxial data equally well for diverse mouse models \cite{bellini2017comparison}, and has been independently validated as the best of multiple commonly used models \cite{schroeder2018predictive}, it is yet phenomenological. It consists of a neo-Hookean term that is meant to capture the elastin-dominated isotropic responses plus four Fung-exponential terms that are meant to capture the collagen-dominated anisotropic responses; these four fiber families were motivated by multiphoton microscopy, but used mainly because of the important and yet not well understood role of extensive cross-links amongst the different fiber families. Because the relation is phenomenological, one cannot ascribe physical meaning to the material parameters. Yet, of the eight parameters the one that might be thought to be most descriptive is the single neo-Hookean parameter. Its values range from about 21 to 32 kPa for many WT DTAs, whereas its value was found to be 25.0 kPa for the hBAC-mWT, 29 kPa for the $\textit{Fbn1}^{\text{mgR/mgR}}$, 17.4 for the $\textit{Fbln5}^{\text{-/-}}$, and 10 kPa for the hBAC-null \cite{ferruzzi2015decreased,Jiao2017,bellini2016differential}. Hence, although there are some trends in its decrease with decreasing elastic fiber integrity, such is not definitive and there are no trends for the Fung-exponential terms. Clearly, there was a need for another approach for classification.


In this study, we demonstrated the capability of the proposed G2\textPhi net in capturing the genotype-phenotype relationship of soft tissues -- specifically, the genotype-dependent stress-stretch relations for murine aortas. We showed that given measurements of stress response under limited deformation states, G2\textPhi net accurately reconstructs the entire stress-stretch relationship and simultaneously predicts the genotype of the specimen from which the stress data derived. We introduced ensemble learning and mixup regularization to better handle cases with limited data.

G2\textPhi net directly provides a functional expression for a parameterized constitutive relation (Eq.~\ref{eqn:BDTN}) based on the neural operator architecture. By identifying the minimal necessary parameters through an encoder-decoder architecture, G2\textPhi net takes sample variation into account and formulates the sample feature $\bm{\eta}$ with a limited dimension, which together with the injected genotype feature $\bm{\zeta}$ composes the ``material parameters'' of the learned constitutive model. This formulation is formally similar to the classical approach of constitutive modeling by analytical expressions, hence endowing our method with generalizability and transferability across different specimens in multiple material classes. Notably, we do not refer to the existing constitutive models during the entire workflow, including data preprocessing. This provides an end-to-end deep learning approach for constitutive modeling that bypasses existing analytical constitutive models.

While we focused on the application of G2\textPhi net in the biomechanics of aortas, we point out that the utility of G2\textPhi net is general. The mechanics of aortas is not distinctly different from most other load-bearing soft tissues. G2\textPhi net is built on the general theory of continuum solid mechanics, relating deformation states, stress states, and genotypes as material classes. In view of this general theoretical basis, G2\textPhi net will be capable of modeling the biomechanical behavior of other soft biological tissues and relating it to genetic mutations that have biomechanical consequences. More broadly, similar methodology may be applied to modeling even a broader class of solid materials where material classification matters.

We emphasize that, to obtain classification results with high accuracy, the data needs to be sufficiently discriminative to render the genotypes identifiable. In Section S3 in SI (Figs. S2 and S3), we include additional results on model performance for an expanded dataset with 33 specimens from five genotypes, where the additional genotype with 5 specimens comes from the aforementioned hBAC-mWT mouse, the aortas of which contain 115\% of normal elastin. The microstructure of this genotype yields mechanical properties similar to WT, hence the classification accuracy is worse than the aformentioned results for four genotypes.

Despite the effectiveness of G2\textPhi net, there remain questions. First, our method provides point estimations. To quantify uncertainty associated with our predictions, we could utilize recent work \cite{meng2021learning} regarding generative adversarial networks (GANs) for learning a functional prior. In addition, we sought to correlate the constitutive relations with genotype herein. Future studies could focus on building a comprehensive deep learning procedure for capturing the genotype-microstructure-mechanics relationship for mouse models. To this end, an important recent study \cite{holzapfel2021predictive} demonstrated the feasibility of building a machine learning model to estimate material parameters using images representing the microstructural information from arteries.

\section{Materials and Methods} 

\subsection{Animal models}

Data from previously published studies from five groups of mice were re-analyzed herein, all showing different degrees of elastopathy
(\cite{pereira1999pathogenetic,yanagisawa2002fibulin,hirano2007functional}): 
\begin{enumerate}
    \item Wild-type C57BL/6J control mice \cite{Spronck2021},
    \item $\textit{Eln}^{\text{-/-}}$ mice in which elastin production was rescued through introduction of human elastin through a bacterial artificial chromosome (hBAC-mNull), resulting in $\sim 30\%$ of normal elastin expression \cite{Jiao2017},
    \item $\textit{Eln}^{\text{+/+}}$ mice in which elastin production was amplified to $\sim 115\%$ of the normal level through introduction of human elastin (hBAC-mWT) \cite{Jiao2017},
    \item $\textit{Fbn1}^{\text{mgR/mgR}}$ mice, in which expression of the elastin-associated glycoprotein fibrillin-1 is expressed at 15-25\% of its normal expression level \cite{Korneva2019}, and
    \item $\textit{Fbln5}^{\text{-/-}}$ mice, which lack 100\% of the elastin-associated glycoprotein fibulin-5 and biomechanically mimic human arterial aging phenotype \cite{ferruzzi2015decreased, Spronck2020}.
\end{enumerate}
Note that the third type (hBAC-mWT) are used only in results in Section S3 (Figs. S2 and S3) in SI. All mice were euthanized with an intraperitoneal injection of pentobarbital sodium and phenytoin sodium (Beuthanasia-D; 150 mg/kg), after which the DTA was carefully dissected. All animal protocols were approved by the Yale University Institutional Animal Care and Use Committee and conformed to the current Federal guidelines.

\subsection{Biaxial Testing}

DTAs were cleaned of loose perivascular tissue and their side branches ligated. Aortas were then mounted on glass pipets in a custom computer-controlled biaxial testing device and submerged in Hank's Balanced Salt Solution. They were brought to their approximate in vivo axial stretch, determined as the axial stretch for which a given pressure fluctuation from 60 to 140 mmHg led to minimal fluctuation of the measured axial force. Arteries were then equilibrated by applying a pulsatile pressure from ~80-120 mmHg for 15 minutes at this in vivo axial stretch. Finally, arteries were precondidtioned using four inflation-deflation cycles from 10 to 140 mmHg.

We previously showed that combined data sets from multiple pressure-diameter and axial force-length protocols provide information sufficient for robust parameter estimation when using traditional nonlinear constitutive relations in terms of mechanical stress and strain \cite{ferruzzi2013biomechanical}. Arteries were hence subjected to seven protocols: cyclic pressurization from 10 to 140 mmHg while axial length was separately held fixed at its \textit{in vivo} value and ±5\% of this value as well as cyclic axial loading from 0 to $f_{\text{max}}$ while the pressure was separately held fixed at 10, 60, 100, or 140 mmHg. Note that $f_{\text{max}}$ was defined as the maximum value achieved during pressurization to 140 mmHg while the vessel was held at 1.05 times the \textit{in vivo} length. Using standard equations \cite{Humphrey2002a}, biaxial data, in terms of pressure-diameter and axial force-length, were converted to circumferential and axial Cauchy stress-stretch data that were used in the deep learning algorithm.

\subsection{Data Preprocessing}

Before applying G2\textPhi net, we needed to preprocess the raw, unstructured stress-stretch data to obtain structured data on a $m\times m$ ($m=31$ herein) uniform grid of stretch states in $(\lambda_z,\lambda_\theta)\in[1.00,1.65]^2$. To do so, we used a simple fully connected neural network to approximate the stress-stretch relationship, where the stretch state $(\lambda_z,\lambda_\theta)$ serves as the input and strain energy density $w$ serves as the direct output. Stress responses $(\sigma_z,\sigma_\theta)$ are then calculated through automatic differentiation. The neural network is trained by matching the predicted stress to the raw data. To improve the quality of the preprocessed stress data, consider two physical constraints: (1) convexity of the strain energy density $w$, which is a fundamental physical principle; (2) convexity of the stress components as functions of stretch states, which we impose for large stretches (herein defined by $\text{max}\{\lambda_z,\lambda_\theta\}>1.45$) given that the stress-stretch curve is qualitatively J-shaped for relatively large stretches. These two constraints are incorporated into the loss function as penalty terms~\cite{liu2020generic}.

\subsection{Architecture of G2\textPhi net}

G2\textPhi net is shown in  Fig.~\ref{fig:model}; it is based on three neural networks, namely a branch encoder (BE), a branch decoder (BD), and a trunk net (TN). BE is a convolutional neural network that takes stress-strain data as input, in the form of $\{(\sigma_z^{(i,j)},\sigma_\theta^{(i,j)})\}_{i,j=1}^{m,m}$, the stress values at $m\times m$ structured grid points ($m=31$ in this work) of stretch $\{(\lambda_z^{(i,j)},\lambda_\theta^{(i,j)})\}_{i,j=1}^{m,m}$. Since the choices of stretches $\{(\lambda_z^{(i,j)},\lambda_\theta^{(i,j)})\}_{i,j=1}^{m,m}$ are common for all data in this work, they are neglected from the input of the branch encoder. The branch encoder compresses the information on stress into the low-dimensional latent space representation of the sample feature:
\begin{equation}
\label{eqn:BE}
\bm{\eta}=\{\eta^{(i)}\}_{i=1}^{d_{\bm{\eta}}}=\mathcal{NN}_{\bm{\xi}^\text{BE}}(\{(\sigma_z^{(i,j)},\sigma_\theta^{(i,j)})\}_{i,j=1}^{m,m}),
\end{equation}
where $\bm{\xi}^\text{BE}$ represents the trainable parameters of the branch encoder and $2\leq d_{\bm{\eta}}\leq 6$.
BD and TN are both fully connected neural networks. They serve together as an approximation of the stress-stretch relationship parameterized by the sample feature $\bm{\eta}=\{\eta^{(i)}\}_{i=1}^{d_{\bm{\eta}}}$ and class feature $\bm{\zeta}=\{\zeta^{(i)}\}_{i=1}^{d_{\bm{\zeta}}}$, where $d_{\bm{\zeta}}=4$ is the number of relevant genotypes in our problem. The class feature is the one-hot representation of the genotype corresponding to the aorta sample. BD maps the combined representation of class and sample feature into two $p$-dimensional outputs:
\begin{equation}
\label{eqn:BD}
    (\bm{b}_\theta,\bm{b}_z)=(\{b_\theta^{(i)}\}_{i=1}^{p},\{b_z^{(i)}\}_{i=1}^{p})=\mathcal{NN}_{\bm{\xi}^\text{BD}}(\bm{\zeta},\bm{\eta}),
\end{equation}
while TN maps an arbitrary stretch state $(\lambda_\theta,\lambda_z)$ into other two $p$-dimensional outputs:
\begin{equation}
\label{eqn:TN}
(\bm{t}_\theta,\bm{t}_z)=(\{t_\theta^{(i)}\}_{i=1}^{p},\{t_z^{(i)}\}_{i=1}^{p})=\mathcal{NN}_{\bm{\xi}^\text{TN}}(\lambda_\theta,\lambda_z),
\end{equation}
where $\bm{\xi}^\text{BD}$ and $\bm{\xi}^\text{TN}$ are the trainable parameters of BD and TN, respectively. Finally, the stress prediction $(\hat{\sigma}_\theta,\hat{\sigma}_z)$ is calculated by
\begin{align}
\label{eqn:BDTN1}
\hat{\sigma}_\theta=\bm{b}_\theta\cdot\bm{t}_\theta+b_{0\theta}=\sum_{i=1}^pb_\theta^{(i)}t_\theta^{(i)}+b_{0\theta},\\
\label{eqn:BDTN2}
\hat{\sigma}_z=\bm{b}_z\cdot\bm{t}_z+b_{0z}=\sum_{i=1}^pb_z^{(i)}t_z^{(i)}+b_{0z},
\end{align}
where $b_{0\theta}$ and $b_{0z}$ are additional trainable parameters as biases. For simplicity of notation, these two trainable parameters are incorporated into $\bm{\xi}^\text{TN}$ hereafter. Combining the expressions for BD and TN in Eqs. \ref{eqn:BD}-\ref{eqn:BDTN2}, we may write the neural network approximation of the stress-stretch relationship parameterized by class and sample feature as Eq.~\ref{eqn:BDTN} in the main text.

\subsection{Mixup Regularization}

The genotype/class feature $\bm{\zeta}$ is represented by one-hot vectors with dimension $d_{\bm{\zeta}}$. During the learning stage, BD only receives inputs in which the class features are 0-1 binary values; it its thus not well informed of the output behavior for intermediate values of $\bm{\zeta}$, resulting in a possibly non-smooth loss landscape. Such non-smoothness can be detrimental to the succeeding inference stage since the estimated $\bm{\zeta}$ are updated with gradient-based optimizers in deep learning. To regularize the branch decoder to ensure smooth predictions for intermediate values of $\bm{\zeta}$, we adopt the mixup technique for the training data~\cite{zhang2017mixup}. This technique produces synthetic training data by linearly combining features and labels of existing data. Technically, with $n_0$ samples of real data with stress $\{(\sigma_z^{(i,j;k)},\sigma_\theta^{(i,j;k)})\}_{i,j=1}^{m,m}$ and genotype $\bm{\zeta}^{(k)}$ ($k\in\{1,2,...,n_0\}$ indicates the specific sample), we generate the stress and genotype of a synthetic data point according to
\begin{align}
    \label{eqn:mixup_sigma}
    \{(\sigma_z^{(i,j;\text{syn})},\sigma_\theta^{(i,j;\text{syn})})\}_{i,j=1}^{m,m} &= \sum_{k=1}^{n_0}\alpha_k \{(\sigma_z^{(i,j;k)},\sigma_\theta^{(i,j;k)})\}_{i,j=1}^{m,m}\\
    \label{eqn:mixup_genotype}
    \bm{\zeta}^{(\text{syn})} &= \sum_{k=1}^{n_0}\alpha_k\bm{\zeta}^{(k)},
\end{align}
where $\alpha_k$ ($k\in\{1,2,...,n_0\}$) are random, nonnegative weights for different real samples satisfying $\sum_{i=1}^{n_0}\alpha_k=1$. To generate each synthetic data point, we need to choose $n_0$ real samples from different genotypes. After generating the synthetic data, we train G2\textPhi net with both real and synthetic data, which helps to regularize the behavior of the learned constitutive relation (Eq.~\ref{eqn:BDTN}) for intermediate values of $\bm{\zeta}$ (e.g., $\bm{\zeta}=[0.1,0.3,0.0,0.6]^\text{T}$).

\subsection{Constraining the behavior of latent space}

Reconstruction of stress-stretch relationship and classification of genotype are achieved by updating the inputs of BD to fit the provided stress-stretch data pairs in the inference stage. A reliable prediction can be achieved only when the values of $(\bm{\zeta},\bm{\eta})$ are similar in the learning and inference stages. The class feature $\bm{\zeta}$ is naturally constrained to be nonnegative and sum to 1, so that we need minimum efforts to ensure the similarity of $\bm{\eta}$. For the sample feature $\bm{\eta}$, however, we need several techniques to guide its behavior. We consider the following guidelines for ensuring the similarity of $(\bm{\zeta},\bm{\eta})$ in the learning and inference stages:
\begin{itemize}
  \item Enforce the (sample) mean and variance of $\bm{\eta}$ to be close to $0$ and $1$, respectively, in the learning stage. This helps to ensure that the sample feature $\bm{\eta}$ distributes properly around the origin of the $d_{\bm{\eta}}$-dimensional space. Technically, this enforcement is achieved by including a penalty term in the loss function for the learning stage.
  \item Initialize $\bm{\zeta}$ to be $[1/d_{\bm\zeta},...,1/d_{\bm\zeta}]^\text{T}$ and $\bm{\eta}$ to be $[0,...,0]^\text{T}$ in the inference stage.
  \item Ensure that $\bm{\eta}$ never evolves to be too far from the origin in the inference stage. Technically, this is enforced by penalizing $|\bm{\eta}|$ in the loss function for the inference stage. This technique is used only for the Setup 2 (raw unstructured data) of the inference stage (Fig.~\ref{fig:inference_U}C ``w/ Reg.'').
\end{itemize}

\subsection{Loss Function}

In the learning stage (Fig.~\ref{fig:model}A), the model adjusts the value of trainable parameters of the three networks ($\bm{\xi}^\text{BE}$, $\bm{\xi}^\text{BD}$ and $\bm{\xi}^\text{TN}$) to minimize the mismatch of stress at the $m\times m$ grid points between the provided data $\{(\sigma_z^{(i,j)},\sigma_\theta^{(i,j)})\}_{i,j=1}^{m,m}$ and the prediction given by
\begin{equation}
    \label{eqn:BDTN_m}
    \{(\hat{\sigma}_z^{(i,j)},\hat{\sigma}_\theta^{(i,j)})\}_{i,j=1}^{m,m}=\{\mathcal{NN}_{\bm{\xi}^\text{BD},\bm{\xi}^\text{TN}}(\lambda_\theta^{(i,j)},\lambda_z^{(i,j)};\bm{\zeta},\bm{\eta})\}_{i,j=1}^{m,m},
\end{equation}
where $\bm{\eta}$ is given by BE (Eq.~\ref{eqn:BE}). Suppose we have $N$ training samples including synthetic data from mixup, then the reconstruction loss is expressed by
\begin{equation}
    \label{eqn:L1_rec}
    \mathcal{L}_\text{rec} = \frac{1}{m^2N}\sum_{k=1}^{N}v_k\sum_{i,j=1}^{m,m}\Bigg[\frac{(\hat{\sigma}_\theta^{(i,j;k)}-\sigma_\theta^{(i,j;k)})^2}{|\sigma_\theta^{(i,j;k)}+\varepsilon|}+\frac{(\hat{\sigma}_z^{(i,j;k)}-\sigma_z^{(i,j;k)})^2}{|\sigma_z^{(i,j;k)}+\varepsilon|}\Bigg],
\end{equation}
where $\varepsilon$ is a small number to avoid singular results when the true stress is zero, and $v_k$ is the weight for the $k$th training sample. We define this $v_k$ because we need different weights for real data and synthetic data generated by mixup. The loss shown in Fig.~\ref{fig:learning}B is exactly $\mathcal{L}_\text{rec}$. The regularization term of the latent space is
\begin{equation}
    \label{eqn:L1_reg}
    \mathcal{L}_\text{reg} = \frac{1}{2}\Big[\bm{\mu}^\text{T}\bm{\mu}+\text{tr}(\bm{\Sigma})-d_{\bm{\eta}}-\log|\bm{\Sigma}|\Big],
\end{equation}
where $\bm{\mu}$ is the mean ($d_{\bm{\eta}}$-dimensional vector) and $\bm{\Sigma}$ is the covariance matrix ($d_{\bm{\eta}}\times d_{\bm{\eta}}$) of the sample feature $\bm{\eta}$. By minimizing $\mathcal{L}_\text{reg}$, $\bm{\mu}$ tends to be a zero vector and $\bm{\Sigma}$ tends to be the identity matrix, hence enforcing the sample mean and variance of $\bm{\eta}$ to be close to 0 and 1, respectively. The total loss in the learning stage is
\begin{equation}
    \label{eqn:L1}
    \mathcal{L}_\text{learn} = w_\text{rec}\mathcal{L}_\text{rec}+w_\text{reg}\mathcal{L}_\text{reg},
\end{equation}
where $w_\text{rec}$ and $w_\text{reg}$ are weights for two loss terms. $\mathcal{L}_\text{learn}$ is minimized by updating $\bm{\xi}^\text{BE}$ (incorporated through $\bm{\eta}$ in Eq.~\ref{eqn:BDTN_m} and $\bm{\mu}$ and $\bm{\Sigma}$ in Eq.~\ref{eqn:L1_reg}), $\bm{\xi}^\text{BD}$ and $\bm{\xi}^\text{TN}$ (incorporated through $\{(\hat{\sigma}_z^{(i,j)},\hat{\sigma}_\theta^{(i,j)})\}_{i,j=1}^{m,m}$ in Eq.~\ref{eqn:BDTN_m}).

In the inference stage (Fig.~\ref{fig:model}B) (Eq. \ref{eqn:BDTN}), G2\textPhi net is provided with limited, (possibly) unstructured measurements of stress $\{(\sigma_\theta^{(i)},\sigma_z^{(i)})\}_{i=1}^{m'}$ at $m'$ stretch states $\{(\lambda_\theta^{(i)},\lambda_z^{(i)})\}_{i=1}^{m'}$ from an unseen sample. It seeks to minimize the mismatch between stress data and prediction, by updating the class feature $\bm{\zeta}$ and sample feature $\bm{\eta}$ through epochs. The fitting loss is hence calculated by
\begin{equation}
    \label{eqn:L2_fit}
    \mathcal{L}_\text{fit} = \frac{1}{m'}\sum_{i=1}^{m'}\Bigg[\frac{(\hat{\sigma}_\theta^{(i)}-\sigma_\theta^{(i)})^2}{|\sigma_\theta^{(i)}+\varepsilon|}+\frac{(\hat{\sigma}_z^{(i)}-\sigma_z^{(i)})^2}{|\sigma_z^{(i)}+\varepsilon|}\Bigg],
\end{equation}
where $\varepsilon$ is a small number to avoid singular results when the true stress is zero. The regularization term of the latent space is $\mathcal{L}_\text{reg2}=|\bm{\eta}|^4$ (used for Setup 2 with raw unstructured data only; see Figs.~\ref{fig:inference_U}B-C). Hence, the total loss in the inference stage is
\begin{equation}
    \label{eqn:L2}
    \mathcal{L}_\text{infer} = w_\text{fit}\mathcal{L}_\text{fit}+w_\text{reg2}\mathcal{L}_\text{reg2},
\end{equation}
where we adopt the subscript ``reg2'' to distinguish this loss term from that defined for the learning stage in Eq.~\ref{eqn:L1_reg}. $\mathcal{L}_\text{infer}$ is minimized by updating $\bm{\eta}$ and $\bm{\zeta}$ only, seeking to find the best-fit reconstruction of the stress-stretch relationship and identifying the source genotype of the data. In this process, $\bm{\zeta}$ is subject to two constraints: (1) all components are nonnegative; (2) all components sum to 1. After the completion of the iterative inference stage, the updated $\bm{\zeta}$ predicts the genotype according to its largest component, and BD and TN in Eq.~\ref{eqn:BDTN} (with the updated $\bm{\eta}$ and $\bm{\zeta}$) together serve as the learned constitutive relation.

\subsection{Technical Details}

We summarize additional technical details of G2\textPhi net in Section S2 in SI.

\section*{Acknowledgment}
This work was supported, in part, by a grant from the NIH (U01 HL142518). B.S. was supported by the European Union’s Horizon 2020 research and innovation program (No. 793805).

\bibliographystyle{elsarticle-num-names}
\bibliography{reference}


\newpage
\section*{Supporting Information}
\subsection*{Section S1. Results of Biaxial Tests for All Four Genotypes}

\begin{figure}[!ht]
  \centering
  \includegraphics[width=0.7\textwidth]{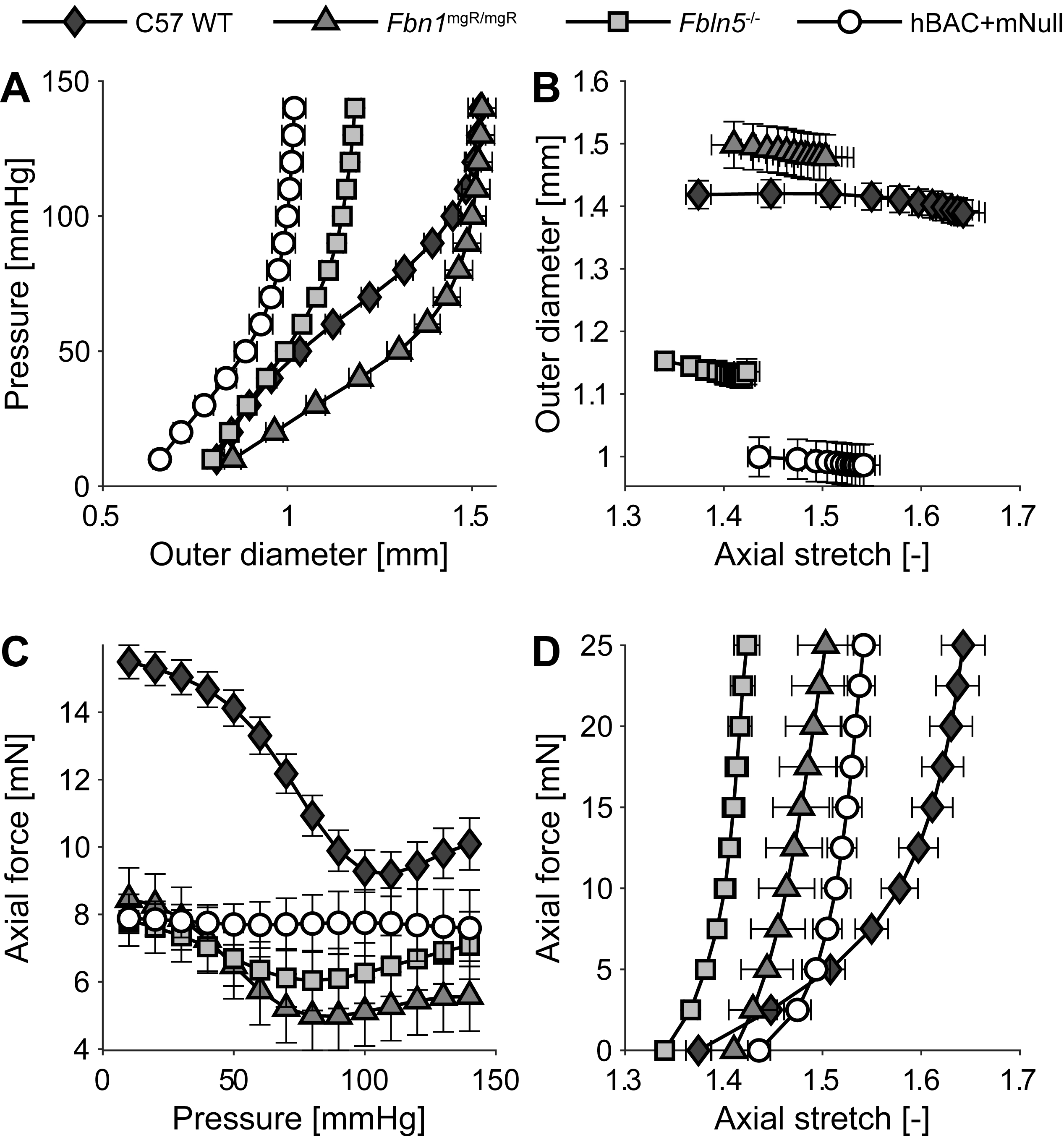}
  \captionsetup{labelformat=empty}
  \caption{Figure S1: \textbf{Biomechanical data of aortas samples from mice with all four genotypes}. Data were obtained through biaxial tests using a seven-step protocol, including three inflation cycles at axial stretch ($\lambda_z$) levels of 95\%, 100\%, and 105\% of the \textit{in vivo} axial stretch ($\lambda_{z\text{,iv}}$) and four extension cycles at pressures ($P$) of 10, 60, 100, and 140 mmHg (C and D). Whiskers indicate standard error ($n=8, 8, 5, 7$ for the four genotypes from left to right of the legend).} 
  \label{figS:experiment}
\end{figure}

\subsection*{Section S2. Technical Details of G2\textPhi net}

G2\textPhi net processes stress data by their log-transformed value defined in Eq.~\ref{eqn:stress_norm}. To further ensure that the neural network deals with stress data that roughly locate within $[-1,1]$, the log-transformed value of stress is divided by $5.0$ before feeding into G2\textPhi net.

BE consists of four convolutional layers each with 64 output channels, with kernel sizes 5, 3, 3, 3 and strides 2, 2, 1, 1, respectively. The output after convolutional layers is then flattened and fed into three dense layers (with sizes $[256, 64, 16, d_{\bm{\eta}}]$). BD consists of four dense layers (with sizes $[d_{\bm{\zeta}}+d_{\bm{\eta}}, 48, 48, 48, 2p]$ where $p=128$). TN consists of four dense layers (with sizes $[2, 48, 48, 48, 2p]$ where $p=128$). Among the $2p$ components in the outputs of BD and TN, $p$ components are used to calculate $\hat{\sigma}_\theta$ and $\hat{\sigma}_z$, respectively (see Eqs.~\ref{eqn:BD}-\ref{eqn:BDTN2}). BE and BD use ReLU activation function for all layers except the final layers (no activation), while TN uses Tanh activation function for all layers.

In the learning stage, the learning rate is initially 0.001 and halves every 5000 epochs. We set $\varepsilon=0.02$, $w_\text{rec}=1.0$, $w_\text{reg}=0.01$; $v_k=0.01$ if the $k$th data is synthetic and $v_k=1$ otherwise. To balance the four genotypes with imbalanced dataset (8, 8, 5, and 7 samples, respectively), we oversample certain data points from the minority classes (i.e., the genotype with 5 and 7 samples) to make sure that the data is balanced. We generate 56 synthetic samples by mixup. Both the mixup and oversampling are random and changes every epoch. The number of copies saved in the training process is $K_2=3$, corresponding to 16K, 18K, and 20K epochs.

In the inference stage, the learning rate is 0.1. We set $\varepsilon=0.02$, $w_\text{fit}=1.0$, and $w_\text{reg2}=10^{-5}$ (for Setup 2 with raw unstructured data only). Sample feature $\bm{\eta}$ is defined as trainable variable and directly fed into BD. To define the class feature, we firstly define logits $\bm{\zeta}'\in\mathbb{R}^{d_{\bm{\zeta}}}$, and then feed $\bm{\zeta}'$ into a softmax layer to calculate the class feature $\bm{\zeta}=\text{Softmax}(\bm{\zeta}')$. In this way, we ensure that the class feature $\bm{\zeta}$ is nonnegative and sums to 1.

We conduct five-fold cross-validation in our study. Specifically, as we have 8, 8, 5 and 7 samples for the four genotypes, we keep one sample from every genotype as test sample each time, resulting in 24 training samples and 4 testing samples.

\subsection*{Section S3. Model Performance for Genotypes with Similar Biomechanical Properties}

\begin{figure}[!ht]
  \centering
  \includegraphics[width=\textwidth]{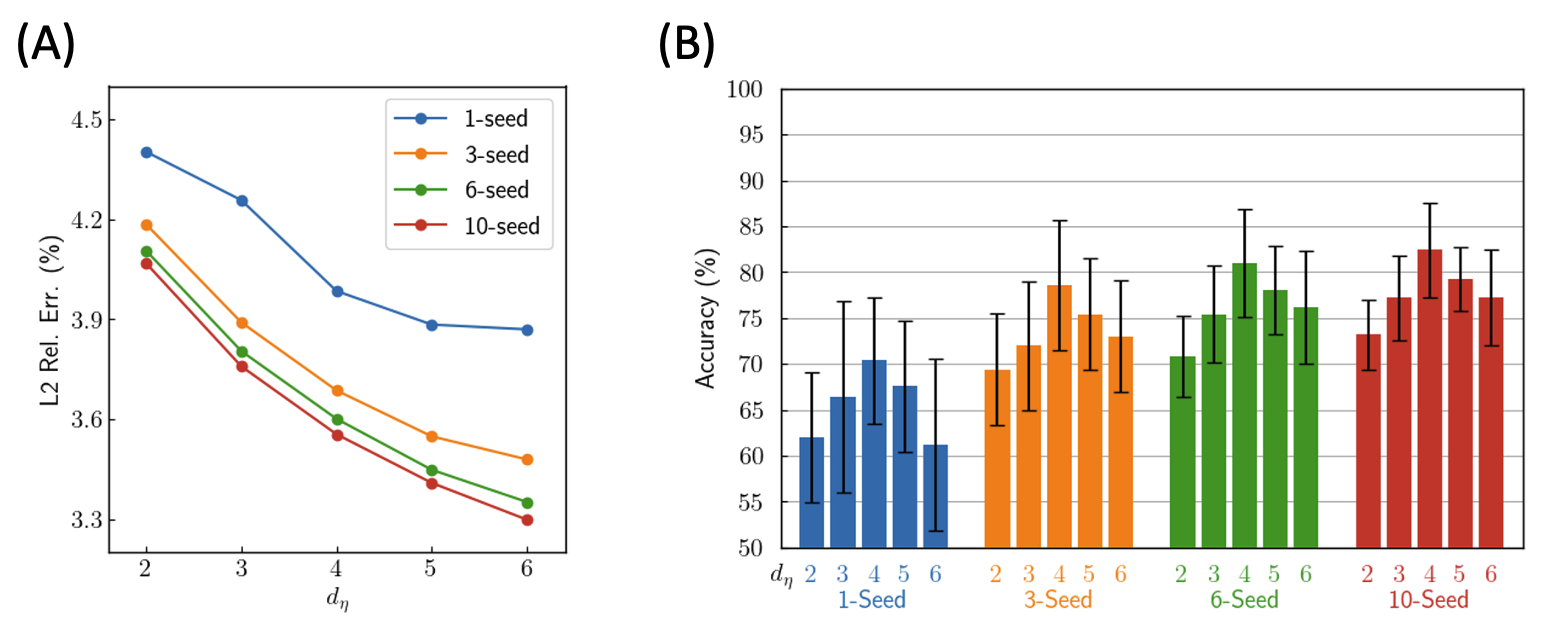}
  \captionsetup{labelformat=empty}
  \caption{Figure S2: \textbf{Reconstruction and Classification Results of Sample Inference based on Structured Data for Expanded Dataset with Five Genotypes}. Samples with five genotypes, where two of them (WT and hBAC-mWT) share similar microstructure and hence exhibit similar mechanical properties, are involved. (A) $L_2$ relative error of the predicted normalized stress $\tilde{\sigma}_i$ with different sizes of ensemble (1-, 3-, 6- and 10-seed) and different dimensions of sample feature ($d_{\bm\eta}\in\{2,3,4,5,6\}$). Each dot is an average value over 20 runs with different choices of random seeds. (B) Classification accuracy for different sizes of ensemble and different dimensions of sample feature. Each bar is the average value over 20 runs with different choices of random seeds.}
  \label{figS:S}
\end{figure}

\begin{figure}[!ht]
  \centering
  \includegraphics[width=\textwidth]{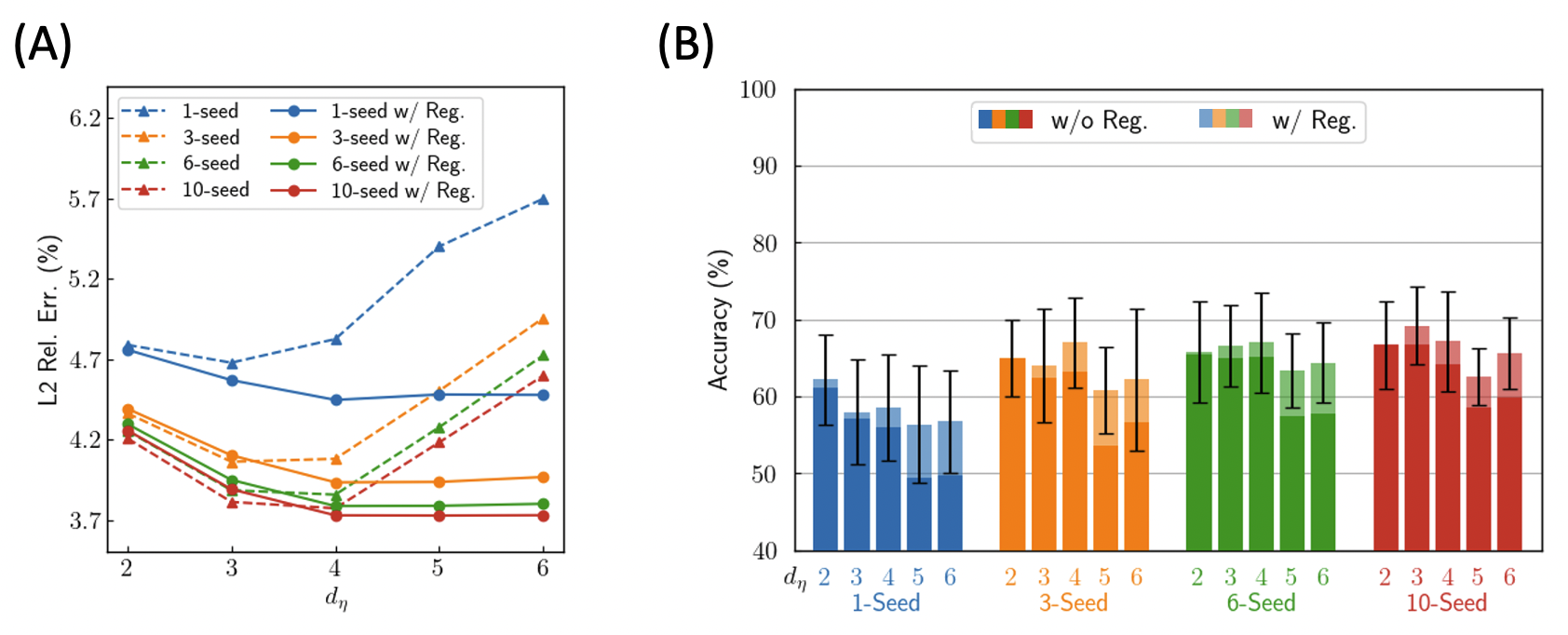}
  \captionsetup{labelformat=empty}
  \caption{Figure S3: \textbf{Reconstruction and Classification Results of Sample Inference based on Unstructured Data for Expanded Dataset with Five Genotypes}. Samples with five genotypes, where two of them (WT and hBAC-mWT) share similar microstructure and hence exhibit similar mechanical properties, are involved. (A) $L_2$ relative error of the predicted normalized stress $\tilde{\sigma}_i$ with different sizes of ensemble (1-, 3-, 6- and 10-seed), different dimensions ($d_{\bm\eta}\in\{2,3,4,5,6\}$) and regularization for the sample feature (without/with regularization). Each dot is the average value over 20 runs with different choices of random seeds. (B) Classification accuracy for different sizes of ensemble, different dimensions and different regularization setups of sample feature. Each bar is the average value over 20 runs with different choices of random seeds.}
  \label{figS:U}
\end{figure}

\end{document}